\documentclass[a4paper]{jpconf}
\usepackage{graphicx}

\begin{document}
%\linenumbers
\title{Commissioning of the ATLAS Level-1 Trigger with Cosmic Rays}

\author{Thilo Pauly\footnote{on behalf of the ATLAS TDAQ Level-1 Trigger group: R.~Achenbach, 
P.~Adragna, 
G.~Aielli, 
A.~Aloisio, 
M.G.~Alviggi, 
V.~Andrei, 
S.~Antonelli, 
S.~Ask, 
O.~Bahat Treidel, 
B.M.~Barnett
% ...
}}

\address{CERN, Switzerland}

\ead{thilo.pauly@cern.ch}

\begin{abstract}
The ATLAS detector at CERN's Large Hadron Collider will be exposed to proton-proton collisions from beams crossing at 40~MHz. A three-level trigger system was designed to select potentially interesting events and reduce the incoming rate to 100-200~Hz. The first trigger level (LVL1) is implemented in custom-built electronics, the second and third trigger levels are realized in software. Based on calorimeter information and hits in dedicated muon-trigger detectors, the LVL1 decision is made by the central-trigger processor yielding an output rate of less than 100~kHz. The allowed latency for the trigger decision at this stage is less than 2.5 microseconds. 
Installation of the final LVL1 trigger system at the ATLAS site is in full swing, to be completed later this year. We present a status report of the main components of the first-level trigger and the in-situ commissioning of the full trigger chain with cosmic-ray muons.
\end{abstract}

\section{ATLAS Level-1 Trigger System}
\begin{figure}[h]
\begin{minipage}{14pc}
\includegraphics[width=14pc]{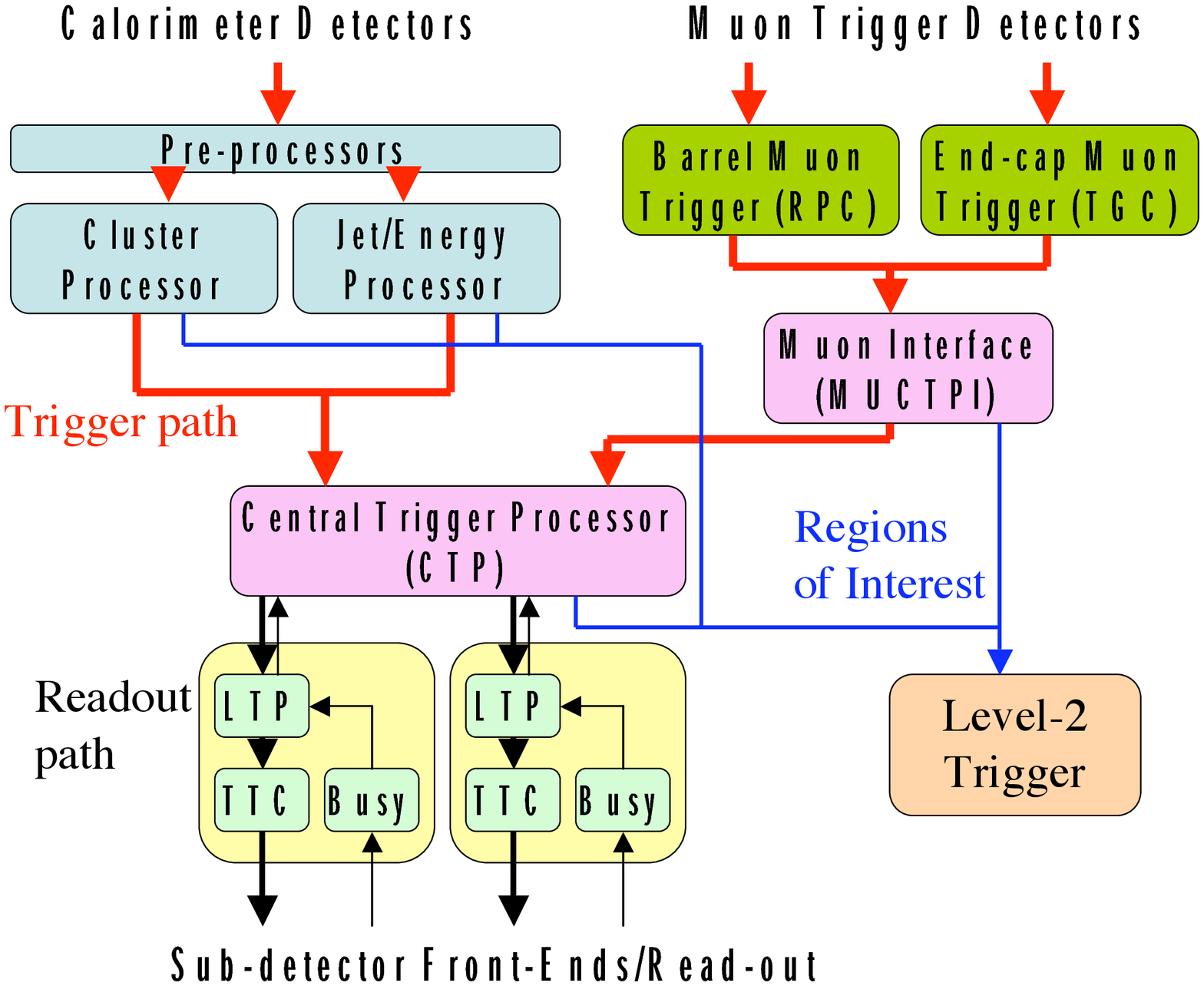}
\caption{\label{fig:L1}Figure caption for first of two sided figures.}
\end{minipage}\hspace{2pc}%
\begin{minipage}{14pc}
\includegraphics[width=14pc]{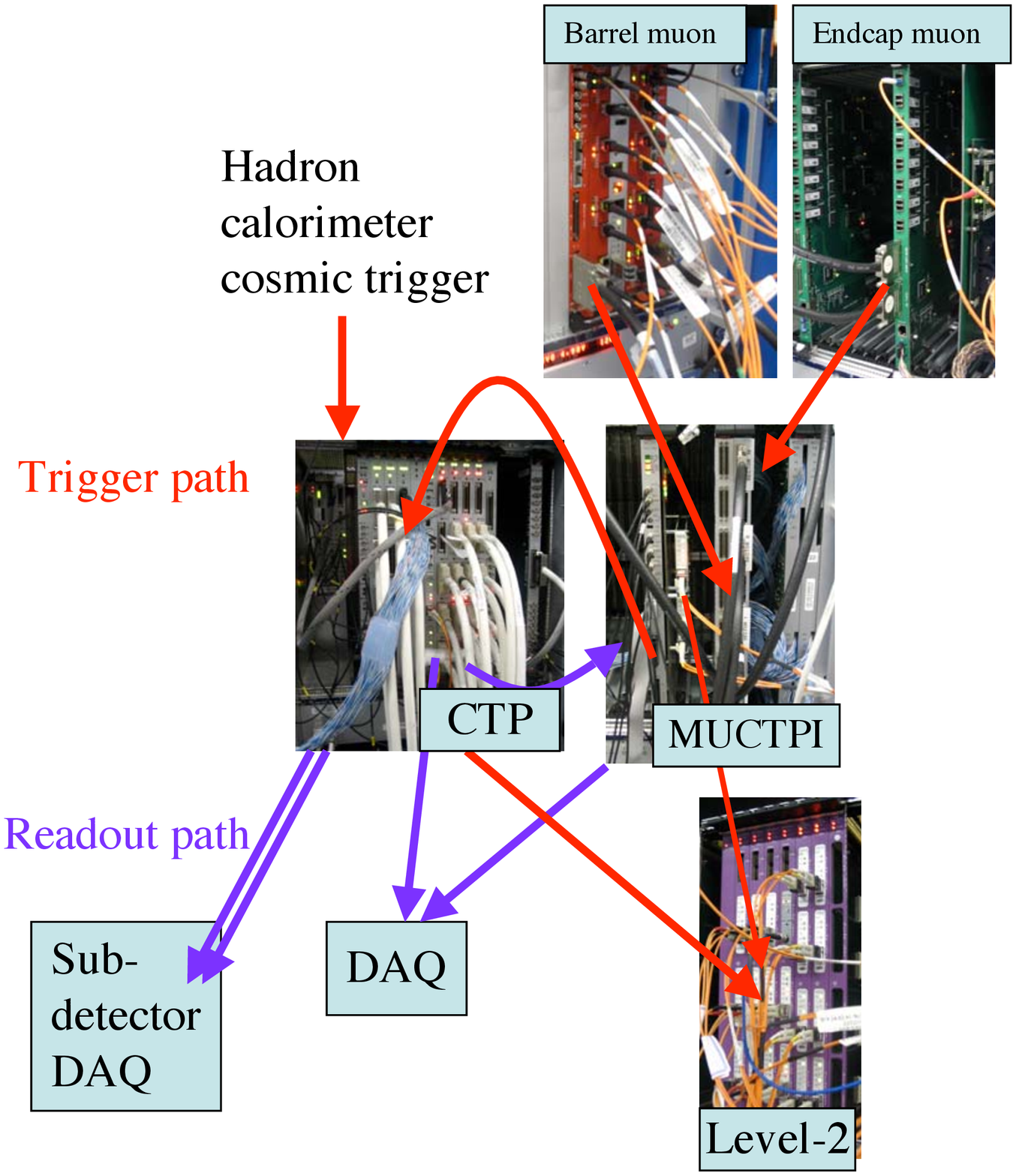}\hspace{2pc}%
\caption{\label{fig:M3}Figure caption for second of two sided figures.}
\end{minipage} 
\end{figure}

The ATLAS Level-1 trigger \cite{bib:AtlasLevel1} is a system that synchronously processes information from the calorimeter and muon trigger detectors at the bunch crossing frequency of 40~MHz.
It consists of three parts: the calorimeter trigger, the muon trigger, and the central trigger (see Fig.~\ref{fig:L1}.
Trigger information from the calorimeters is processed in the Calorimeter Trigger system, which sends electron/photon, tau/hadron, jet candidates as well as global energy information to the Central Trigger Processor (CTP).
Muon trigger candidate information from the resistive-plate chambers in the barrel region and thin-gap chambers in the end-cap region is sent to the Muon-to-Central-Trigger-Processor-Interface (MUCTPI), which subsequently forms global muon multiplicities for six different transverse momentum ($p_T$) thresholds and sends them to the CTP.
From this information the CTP makes the final Level-1 Accept (L1A) trigger decision according to a programmable trigger menu -- a table of logic conditions that are used to make the final trigger decision.
The L1A is then fanned out to all sub-detectors to initiate readout of the triggered event.
In addition, the Level-1 trigger systems send region-of-interest (RoI) information to the Level-2 trigger and take part in the combined readout of ATLAS.

%-----------------------------------
\section{Commissioning of the ATLAS Level-1 Trigger System}
\label{sectionStatus}
%-----------------------------------
In June 2007 we have performed a cosmic-ray run combining a large slice of ATLAS in-situ with the aim of exercising the full data recording chain, from detector to disk, including the full trigger path.
The Level-1 trigger system used to trigger this data consisted of the barrel muon trigger, the end-cap muon trigger, and a temporary trigger formed from coincidences of signals from the upper and lower part of the hadronic calorimeter.
This Level-1 trigger was used to read out the following sub-detectors: silicon strip tracker, transition radiation tracker, electromagnetic calorimeter, hadronic calorimeter, muon precision chambers, and the Level-1 trigger system.
The readout of the Level-1 trigger system not only included the central trigger and the barrel and end-cap muon trigger systems, all of which took part in forming the trigger decision, but also a fraction of the calorimeter trigger electronics.
A muon algorithm was run online in the Level-2 trigger system, seeded by regions-of-interest from the MUCTPI and CTP.

Most of the chambers of the barrel muon trigger are installed and being commissioned.
The off-detector electronics is in production and demonstrators have been used for commissioning.
During the cosmic-ray run in June 2007 the top sector provided a cosmic-ray trigger to the MUCTPI and CTP through the final trigger chain.
The rate was about 120~Hz.
For the first time the chambers have been operated with the final gas setup, cabling, power system and slow control.
A trigger latency of 1490~ns was measured from the chamber to the output of the CTP, which is well within specifications.
Synchronization tests between RPC layers have been successfully performed.

The chamber installation of the end-cap muon trigger is in full swing.
The final on-detector and off-detector electronics is being installed and commissioned.
For the June 2007 cosmic-ray run, the inner layer of one end-cap muon trigger sector provided a cosmic-ray trigger signal, which was sent to the MUCTPI and CTP through the final trigger chain.
The trigger rate was a few Hz.
The measured trigger latency was 1550~ns from the chamber to the output of the CTP, in agreement with expectations.
The end-cap muon trigger system was integrated in the data-acquisition using final readout equipment.

The MUCTPI and CTP are installed in the underground counting room since January 2006.
While the CTP uses final components and is complete, the MUCTPI still uses demonstrator boards which are close to the final version.
The CTP and MUCTPI are routinely used during cosmic-ray runs and commissioning to provide a Level-1 trigger.

The installation of the Level-1 calorimeter trigger is moving quickly:
about 40~\% of the pre-processor are installed, the jet-energy processor installation is almost complete, the cluster processor installed to about 50~\%, and the complex cabling (analogue and digital) fully installed.
All hardware will be installed by October 2007.
The system is currently being commissioned in vertical slices through the system and tests of integration are ongoing with the data-acquisition and the Level-2 trigger system.
In the June 2007 cosmic-ray run, an initial small test system (pre-processor and readout) took part successfully in the combined data-taking.

Fig.~\ref{fig:M3} shows schematically the setup of the trigger logic used in the June 2007 cosmic-ray run.
The two inlay pictures on the top right show the sector-logic outputs of the barrel and end-cap muon trigger systems, which send muon candidate information at 40~MHz to the MUCTPI.
The MUCTPI handles overlaps and forms muon sums of different transverse-momentum thresholds.
These multiplicities are sent to the CTP.
In parallel, the cosmic trigger from the hadronic calorimeter is fed into the CTP.
In the CTP, the trigger signals are aligned in time and compared against the trigger menu.
This trigger menu was loaded into the CTP at the start of each data-taking session from the trigger database, which is the final mechanism for configuring the trigger menu.
The trigger menu that was used was simple: any muon from the hadronic calorimeter or the muon trigger systems was formed into a L1A.
For each accepted event, the L1A was sent to all the sub-detectors to initiate the readout of their detector, including the Level-1 trigger system.
In addition, the CTP and MUCTPI sent region-of-interest information to the Level-2 trigger system, which seeded online reconstruction and selection algorithms (see \cite{bib:ricardo}).

The data taken with the Level-1 trigger in June 2007 is currently being analyzed and checked for internal consistency.
The data is rich in cosmic-ray muons, confirmed by the precision muon chambers, which see particle tracks for both end-cap and barrel muon triggered events.
From the comparison of the time spectra in the precision chamber it was possible to measure that the trigger signals from the barrel reaches the MUCTPI about 130~ns sooner than from the end-cap muon trigger chambers.
In addition, there was confirmation on cosmic-ray muons from the electromagnetic and hadronic calorimeters, as well as from the transition radiation detector.

\end{document}